\documentclass[twocolumn]{revtex4}          

\def\iso#1#2{\mbox{${}^{#2}{\rm #1}$}}

\def\li#1{\iso{Li}{#1}}
\def\c1#1{\iso{C}{1#1}}
\def\n1#1{\iso{N}{1#1}}
\def\o1#1{\iso{O}{1#1}}

\def\beq{\begin{equation}}
\def\eeq{\end{equation}}
\def\beqar{\begin{eqnarray}}
\def\eeqar{\end{eqnarray}}
\def\pfrac#1#2{\left( \frac{#1}{#2} \right)}

\def\la{\mathrel{\mathpalette\fun <}}

\def\fun#1#2{\lower3.6pt\vbox{\baselineskip0pt\lineskip.9pt
  \ialign{$\mathsurround=0pt#1\hfil##\hfil$\crcr#2\crcr\sim\crcr}}}

\begin{document}

\title{Cosmological Cosmic Rays: \\
Sharpening the Primordial Lithium Problem}

\author{Tijana Prodanovi\'{c}}
\email{prodanvc@if.ns.ac.yu}
\affiliation{Department of Physics, University of Novi Sad \\
Trg Dositeja Obradovi\'{c}a 4, 21000 Novi Sad, Serbia}

\author{Brian D. Fields}
\affiliation{Department of Astronomy, University of Illinois \\
1002 W. Green St., Urbana IL 61801, USA}

\begin{abstract}

Cosmic structure formation leads to large-scale shocked baryonic flows
which are expected to produce a cosmological population of
structure-formation cosmic rays (SFCRs).  
Interactions between SFCRs
and ambient baryons will produce 
lithium isotopes via $\alpha + \alpha \rightarrow \iso{Li}{6,7}$.
This pre-Galactic (but non-primordial)
lithium should contribute to the primordial \li7
measured in halo stars and must be subtracted in order to
arrive to the true observed primordial lithium abundance. 
In this paper we point out that the recent halo star \li6 measurements
can be used to place a strong constraint to the level of such contamination,
because the exclusive astrophysical production of
\li6 is from cosmic-ray interactions. 
We find that the putative \li6 plateau, if due to pre-Galactic cosmic-ray
interactions, implies that SFCR-produced lithium represents
$\rm Li_{SFCR}/Li_{\rm plateau}\approx 15\%$ of the observed elemental Li 
plateau.
Taking the remaining plateau Li to be cosmological \li7,
we find a revised (and slightly worsened) 
discrepancy between the Li observations and Big Bang 
Nucleosynthesis  predictions
by a factor of $\li7_{\rm BBN}/\rm \li7_{\rm plateau} \approx 3.7$. 
Moreover, SFCRs would also contribute to the
extragalactic gamma-ray background (EGRB) through neutral pion production. 
This gamma-ray production is tightly related to the amount of 
lithium produced by the same cosmic rays; the \li6 
plateau limits the
pre-Galactic (high-redshift) SFCR contribution to be at the level of 
$I_{\gamma_{\pi}^{}\rm SFCR}/I_{\rm EGRB}\la 5\%$ of the currently observed 
EGRB. 

\end{abstract}

\maketitle

\section{Introduction}

The observation of the lithium plateau in low-metallicity halo stars 
\cite{spite} indicates pre-Galactic lithium production,
and has long been understood as signature of the 
primordial lithium predicted by the Big Bang Nucleosynthesis (BBN) theory.
However, recent {\it WMAP} results \cite{wmap} together with the BBN theory
predict the primordial lithium abundance 
$(\li7/{\rm H})_{\rm BBN}=3.82^{+0.73}_{-0.60} \times 10^{-10}$ \cite{cyburt} 
that is a factor of $3$ higher than than the observed elemental lithium 
plateau abundance of
$(\rm Li/{\rm H})_{\rm pl}=1.23^{+0.34}_{-0.16} \times 10^{-10}$ \cite{ryan}.
Moreover, any non-primordial but pre-Galactic source of lithium would
act as a ``contaminant'' to the plateau lithium abundance and would have
to be corrected for in order to obtain the true primordial plateau,
which would, consequently, create an even larger discrepancy between theory
and observations and result in a even larger lithium problem. 

A very well-motivated 
candidate for such pre-Galactic source of \li7 would be a 
cosmic-ray population that would originate 
during the 
process of cosmological structure formation.
Specifically, the particles should be
accelerated in the shocks which
inevitably arise from the infall
of bayronic matter onto dark matter potentials
\cite{miniati}.
The composition of these cosmic rays would be primordial, i.e. made only of
protons and alpha particles; their interactions with
ambient baryons produces lithium isotopes via
$\alpha + \alpha \rightarrow \iso{Li}{6,7}$
\cite{alphaalpha}. 
Besides \li7, any cosmic ray population would also produce \li6.
Unlike \li7, the {\it only} known astrophysical nucleosynthesis
mechanism for \li6 production is in cosmic-ray interactions
\cite{li6nuke}. 
Thus, if structure formation cosmic rays (SFCRs) are important
pre-Galactic source of lithium, they should also result in \li6 production
\cite{si02}, and a \li6 plateau should also exist at some level in low-metallicity halo stars.

Recent halo star observations indeed indicate
the existence of a \li6 plateau \cite{asplund}. 
These high-sensitivity spectra measure the Li line shape 
precisely enough to obtain an isotope ratio
$\li7/\li6 \approx 0.05$, which corresponds to a plateau
of ${\rm \li6/H} \approx 6 \times 10^{-12}$.
This lies far above the standard BBN level of \li6 production
\cite{thomas}, 
and thus has provoked enormous interest.
Some scenarios for decaying dark matter can allow 
for \li6 production 
\cite[e.g.][]{jedamzik}.
It was also suggested that the \li6 may not be pre-Galactic
but due to {\em in situ} flare production \cite{tatischeff}.
In the present discussion, we will work within the
assumption that there is a \li6 plateau, which 
indicates a pre-Galactic \li6 component, whose origin
is not primordial but astrophysical--i.e., due to accelerated particles.

Since the {\em ratio} of 
\li6 and \li7 production in cosmic-ray interactions depends {\it only}
on their cross sections, the existence of the \li6 plateau can be used to
determine the possible production of pre-Galactic \li7 and constrain the
possible ``contamination'' to the Spite plateau by the SFCR population,
or any other pre-Galactic cosmic-rays \cite{fp,rvo05,rvo06}. 
Such a correction is in addition to--but a logical extension of--the
correction due to Li (and Be and B) synthesis by Galactic cosmic-rays,
which themselves have a small impact on the plateau value
\cite{LiGCR}.
We find below
that SFCRs can then make up to $15\%$ of the observed elemental lithium
 plateau at best. This is in agreement with the findings of \cite{rvo05}
who analyzed the observed \li6 plateau as a result of cosmic-rays
that would originate from Population III stars.

Besides lithium, SFCRs
would also give rise to the gamma ray emission from inverse Compton
scattering of electrons off photon background and from decay of neutral
pions that would result from hadronic interactions 
$pp\rightarrow \pi^0 \rightarrow 2\gamma$ \cite{lw}. 
This would contribute to the observed extragalactic gamma-ray background
\cite{strong}.
It was shown in \cite{fp} that there is a tight connection between
pionic gamma rays and lithium that are produced by a given cosmic ray
population. Thus, by using this connection and our constraint on the
possible \li7 production by the SFCRs, we can also constrain the
level at which this cosmic-ray population could contribute
to the observed extragalactic gamma-ray background (EGRB).
We find below that the SFCRs can in the upper limit contribute at the level of 
$5\%$ to the EGRB. 

The work of \cite{rvo05} has been along similar lines of argument as
those presented here.
In their paper \cite{rvo05} (and the followup \cite{rmvoi}) account for the observed \li6 plateau by
cosmic-ray interactions where they consider cosmic-rays that would
originate from early Population III stars, as opposed to structure formation
cosmic ray population discussed here. In this paper we place even stronger
constraints, but we also draw attention to how the two scenarios of
different cosmic ray populations could be discriminated against.

\section{An Estimate of the \li7 Production by Structure Formation Cosmic Rays}

The ratio at which \li7 and
\li6 are made in cosmic-ray interactions depends only on the ratio of their
production reaction rates, which is a ratio of their production cross 
sections, weighted by the cosmic-ray energy spectrum. 
In the case of SFCRs, the only relevant 
production channel is through fusion reaction 
$\alpha + \alpha \rightarrow ^{6,7}{\rm Li}$. 
With the adopted SFCR spectrum characteristic of strong shocks
the ratio at which \li7 and \li6 are produced through this channel
is 2:1 \cite{fp}.  
If we assume that the observed \li6 plateau abundance of
$\li6/{\rm H}=6.3 \times 10^{-11}$ \cite{asplund} is entirely made
by SFCRs that originate from strong cosmological shocks, this
results in the $\rm Li=\li7+\li6$ production of 
\beqar
\pfrac{\rm Li}{\rm H}_{\rm SFCR} & = & 1.9 \times 10^{-11}    \\
\frac{{\rm Li}_{\rm SFCR}}{{\rm Li}_{\rm plateau}} & = & 0.15
\eeqar

With these estimates we can now revise the existing discrepancy
between \li7 plateau observations and BBN prediction. We correct
the plateau abundance for the possible contamination by SFCRs
\beq
\frac{\li7_{\rm BBN}}{{\rm Li}_{\rm plateau}-\li7_{\rm SFCR}-\li6_{\rm SFCR}}
  = 3.7
\eeq
and find that the magnitude of this discrepancy 
is enlarged by $\sim 25\%$.
We again emphasize that this revised limit assumes
the \li6 is due to SFCRs, but is independent of the
details of the SFCR spectra which all give similar
\li7/\li6 ratios.

Because \li6 is the more fragile isotope, 
it is more susceptible to {\em in situ} stellar
depletion effects which one must always consider.
Ref.~\cite{asplund} use the pre-main-sequence stellar
models of \cite{richard} to estimate the possible impact of
depletion; the resulting corrected \li6 abundances
now show a nonzero rising slope in the \li6
abundance with respect to the metallicity rather than a plateau-like feature.
However, for the purpose of our argument
such a \li6-metallicity trend 
would still constrain the SFCR lithium
yield, because SFCRs should still give rise to a \li6 plateau
that should resurface below some metallicity. 
In this case
the lowest \li6 abundance represents
an upper limit to the \li6 production by SFCRs. 
This scenario 
gives somewhat higher estimates of \li6 abundances in low-metallicity halo
stars and the accompanying SFCR \li7 limit would increase to
${\rm Li}_{\rm SFCR}/{\rm Li}_{\rm plateau} \approx  0.24$.
This propagates to give a true primordial Li plateau abundance
of $\li7_{\rm plateau,true} \approx 9\times 10^{-11}$ and increases
the discrepancy with BBN predictions to the factor of 4.2.

\section{An Estimate of the hadronic SFCR contribution to the EGRB}

In \cite{fp} we have shown and quantified the tight
connection between lithium and hadronic $\gamma$-ray production through
cosmic-ray interactions.
With the assumption that cosmic-ray history in our Galaxy is 
representative of an average star-forming galaxy, this connection
can be expressed as
\beq
\label{eq:gamma/Li}
I_{\gamma_\pi}(E>0)
 = I_{0,i} \frac{Y_{i,\rm obs}} {Y_{i,\odot}}   
\eeq
where $Y_i \equiv n_i/n_b$ measures the abundance
of $i \in \li6,\li7$ per baryon.
The factor $I_{0,i}$ depends on the assumed helium abundance in
cosmic rays and in the local medium and also incorporates the ratio of
flux averaged cross sections for $pp\rightarrow \pi^0 \rightarrow 2\gamma$ and
$\alpha \alpha \rightarrow ^{6,7}{\rm Li}$ production reactions.
For the adopted cosmic-ray spectrum representative of strong shocks and
using \li6 as an indicator, we adopt this prefactor to be
$I_{0,6}=1.86 \times 10^{-5} \rm cm^{-2} \ \ s^{-1} \ \ sr^{-1}$ \cite{fp}.
We note here that $I_{\gamma_\pi}(E>0,t)$ is the total pionic gamma-ray
intensity, i.e. integrated over the entire energy range.
Using the solar \li6 abundance from \cite{ag} and the observed
\li6 plateau abundance of \cite{asplund} we find that SFCRs can at best
produce 
$I_{\gamma_\pi}(E>0) = 7.7 \times 10^{-7} \ \rm cm^{-2} \ s^{-1} \ \ sr^{-1}$
pionic gamma rays over the entire energy range. 

To be able to compare this with observations which have some lower
energy limit one would have to assume something about the history of pionic
gamma-ray production by SFCRs. Since any such assumption would be quite
model dependent we will only try to provide a model-independent upper limit
to the SFCR pionic gamma-ray fraction to the EGRB. We will do this by assuming
that all of the SFCR pionic gamma-rays are created at redshift zero, since
higher redshifts would put more weight on the lower energy part of the pionic
gamma-ray spectrum. For example, for the pionic gamma-ray spectrum
adopted from \cite{ensslin} and a strong-shock cosmic-ray spectrum
we find that if all of the pionic gamma-rays made by SFCRs are taken to 
originate from redshift zero, then 
$I_{\gamma_\pi}(z=0, E>0.1 {\rm GeV})/I_{\gamma_\pi}(z=0, E>0 \rm GeV) = 0.77$,
compared to the case where we assume that they all originate from $z=10$ where
we now get $I_{\gamma_\pi}(z=10, E>0.1 {\rm GeV})/ I_{\gamma_\pi}(z=10, E>0 \rm GeV)=0.12$.
Thus, if we assume that all of the SFCR pionic gamma rays come from $z=0$,
this gives us the upper most limit and
we find that
$I_{\gamma_\pi}(z=0, E>0.1 \rm GeV)
  = 5.9 \times 10^{-7} \ \rm cm^{-2} \ s^{-1} \ \ sr^{-1}$
which is $\approx 5\%$ of the observed EGRB 
$I_{\gamma, obs} (E>0.1 \rm GeV)
  = 1.1 \times 10^{-5} \rm \ cm^{-2} \ s^{-1} \ \ sr^{-1}$ \cite{strong}.

It is important to bear in mind the nature of 
the gamma-ray/lithium connection
encoded in eq. \ref{eq:gamma/Li}.
The common cosmic-ray origin of Li isotopes and
pionic photons links both observables
at any epoch $t$ to the cosmic-ray fluence (integrated flux)
{\em up to that epoch}.
This in turn guarantees that the Li abundances at $t$
are proportional to the $\gamma$-ray intensity at $t$.
Note, however, that any \li6 plateau abundance
must have been produced pre-Galactically, i.e., at high redshift.
Thus, the pionic $\gamma$-rays associated with the plateau \li6
are only those produced by SFCRs {\em at redshifts prior to halo
star formation}.
Any additional, post-halo-star SFCR activity will contribute
(at lower redshifts) to the pionic background, but not to
the halo-star \li6 plateau.
Hence, our pionic limit is only on the high-redshift
EGRB component; a lower redshift SFCR contribution could exist.
With this in mind, we note the following.
(1) The very existence of any \li6 plateau demands 
a pre-Galactic origin, which if astrophysical
would in turn require a rapid and high-redshift 
particle flux from SFCRs (or Pop III supernovae);
these particles must contribute to a pionic $\gamma$-ray background
at some level.
(2) Turning the problem around, if a diffuse, redshifted pionic signature
can be found in the EGRB, this places an {\em upper} limit
on pre-Galactic Li from accelerated particles.
If this limit is near the \li6 plateau, one could even hope to
use the redshift of the pionic feature as an indicator of
the epoch of halo star formation.

\section{Discussion}
\label{sect:discussion}

In this paper we have have placed strong constraints to the level at which
a structure formation population of cosmic rays could contribute to the
pre-Galactic lithium production and thus ``contaminate'' halo-star
measurements of the lithium plateau which should reflect the primordial
\li7 abundance. We find that SFCRs can at most contaminate the Spite
plateau at the level of $15\%$. This in turn makes the discrepancy between
BBN predicted primordial \li7 abundance and the observed plateau even larger.
The two values differ now by the factor of $\approx 3.7$.  
Thus the cosmological \li7 problem is indeed worsened
but only mildly,
if pre-Galactic
cosmic rays are the source of the \li6 plateau.
But it is worth emphasizing that in this scenario
(1) the cosmological \li7 problem does remain, and that
(2) any solution to the problem must account for
the \li7 discrepancy but {\em avoid} \li6 production
at or above the observed plateau.
That is, the mere existence of a \li6 plateau {\em does} 
imply pre-Galactic production but {\em does not} necessarily
demand a primordial origin for \li6.

Moreover, we find that SFCR \li6 production
is accompanied by a high-redshift 
pionic gamma-ray flux which would
in the upper limit make up $5\%$ of the present observed EGRB.
Though this represents only a small fraction to the {\it currently} observed 
EGRB, it should certainly leave an imprint on the new observation of the EGRB
by {\it GLAST} \cite{glast}. Though no physical feature is at present seen
in the EGRB spectrum, greater sensitivity of {\it GLAST} will allow for
many of the currently unresolved sources to become resolved which will
result in a lower EGRB \cite[e.g.][]{stecker}. 
Pionic gamma-rays made in SFCR interactions
represent a true diffuse component of the EGRB and will thus contribute
even more to the new reduced EGRB. A spectral feature in such a
diffuse 
component \cite{fp} could then potentially be resolved and used to
determine the nature of the cosmic-ray population that gave rise to it.
Namely, the position and the shape of this pionic gamma-ray feature(s)  
could discriminate between arising from a SFCR population and/or some other
early cosmic-ray population \cite{rvo05} because of different source 
histories. 
Unresolved sources are expected to contribute most to the current EGRB at the 
lower energy end \cite{pavlidou07}. Resolving these sources will open a 
window for the signature of SFCR gamma-rays to be seen being that larger 
redshifts of origin of SFCRs will result in larger gamma-ray fluxes at
lower energies. 
Detection of a pionic gamma-ray signature in the EGRB from a given
cosmic-ray population would in turn, also discriminate between different 
explanations of the \li6 plateau.

\acknowledgments
We are thankful to Vasiliki Pavlidou for enlightening discussions.
The work of TP is supported in part by the Provincial Secretariat for Science 
and Technological Development, and by the Ministry of Science of the Republic 
of Serbia under project number 141002B.

{}

\end{document}